# Table-top X-ray Ghost Imaging with Ultra-Low Radiation

Ai-Xin Zhang[*], Yu-Hang He[*], Ling-An Wu[†], Li-Ming Chen[†], Bing-Bing Wang

*Institute of Physics, Chinese Academy of Sciences, Beijing 100191, China*
*University of Chinese Academy of Sciences, Beijing 100049, China*

The use of x-ray imaging in medicine and other research is well known. Generally, the image quality is proportional to the total flux, but high photon energy could severely damage the specimen, so how to decrease the radiation dose while maintaining image quality is a fundamental problem. In "ghost" imaging, an image is retrieved from a known patterned illumination field and the total intensity transmitted through the object collected by a bucket detector. Using a table-top x-ray source we have realized ghost imaging of plane and natural objects with ultra-low radiation on the order of single photons. Compared with conventional x-ray imaging, a higher contrast-to-noise ratio is obtained for the same radiation dose. This new technique could greatly reduce radiation damage of biological specimens.

The use of x-ray imaging in clinical medicine, crystallography, and many other research fields is extensive. Perhaps the most well-known application is x-ray radiography, which has a history of over a hundred years and in its various forms is still the most widely used form of diagnosis, based on x-ray absorption, phase contrast, coherent diffraction, and so on. Generally, in order to get a clear image, the minimum illumination flux is required to be $10^5$ photons/pixel [1]. However, it is the high photon energy that can cause radiation damage on an object, especially biological organisms. Patients exposed to excessive radiation could develop skin burns and increased risk of cancer. The radiation damage is proportional to the dose received, so there is a tradeoff between low radiation damage and image quality. In this paper we shall report the realization of high quality x-ray images with ultra-low radiation by means of "ghost" imaging.

Ghost imaging [2–5] (GI) is an indirect imaging technology based on intensity correlation which can retrieve the spatial information of an object with a non-spatial detector, if we know the distribution of the field incident on the object. The non-spatial detector is called a bucket detector (or single-pixel detector), which is used to collect all the light transmitted or reflected from the object. The first GI experiment was performed with entangled photon pairs produced by parametric down-conversion so the field distribution of the object signal beam was inferred from the correlated idler beam, but this requires a spatially resolving detector in the latter beam [2]. In GI experiments with classical thermal light the field distribution at the object may be measured by means of a beamsplitter, again with one beam passing to the object and the other to a reference detector with spatial resolution [3]. Now, the beamsplitter and reference detector may both be dispensed with, being replaced by a spatially patterned beam produced by some artificial means, such as a computer-controlled digital micromirror device (DMD) or spatial light modulator. Variations of correlated imaging are known severally as computational GI [4,5], single-pixel imaging [6], dual photography [7] and structured illumination imaging [8]. The great interest in GI stems from the fact that, compared with conventional photography, images of high resolution beyond the

---

[*] These authors contributed equally to this work and should be considered co-first authors.
[†] Corresponding author. Email: wula@iphy.ac.cn (L.A.W.); lmchen@iphy.ac.cn (L.M.C.).

Rayleigh diffraction limit [9] may be obtained even with poor illumination [10] in a highly scattering medium [11]. Thus, GI has many potential applications ranging from microscopy [12–14] to 3-dimension GI [15] to long distance lidar [16,17].

In theory, GI is applicable to any wavelength, and indeed has been recently demonstrated with x-rays [18–20] and even atoms [21]. Although the experiment with atoms was beautiful, it is difficult to envisage practical applications, as the object would have to be in the same vacuum chamber as the atoms. X-ray GI (XGI) could have immense potential in many fields, but previously the main stumbling block was that no suitable beamsplitter could be found. Yu et al [18] solved the problem by just shuttling the sample in and out of a relatively stable pseudocoherent x-ray beam; Pelliccia et [19] used a crystal to diffract out the object and reference beams, which however suffered from anticorrelated instabilities due to vibrations of the crystal stage. Both groups performed their experiment at a synchrotron radiation facility, and had to add a monochromator in the beamline to reduce the hard x-ray bandwidth. This year XGI was realized for the first time with a rotating anode x-ray diffraction system as the radiation source [20]. In this case, pseudo-thermal light was used, generated by passing the source beam through a monochromator then through a rotating sheet of copy paper; beam splitting was again realized through crystal diffraction.

In our experiment, we use an incoherent table-top x-ray source and XGI pre-record a series of speckle fields to act as the reference signal for XGI. This is similar in principle to computational GI with visible light, but instead of a rotating ground glass plate or DMD which are ineffective for x-rays we use a piece of sandpaper [22–24]. By this means we have succeeded in obtaining high quality XGI images under ultra-low x-ray illuminance, even at quasi-single-photon levels.

The experimental setup is shown in Fig. 1. The x-ray source is a copper anode Incoatec IμS Stand-alone Generator, operating at 40 kV and 800 μA, and emitting radiation with characteristic wavelengths of 0.15 and 0.14 nm (corresponding to photon energies of 8.04 and 8.90 keV, respectively). A copper plate aperture blocks the unwanted beams, passing only the direct output beam which has a fairly uniform square profile. About 27 cm downstream the beam traverses a 20 cm diameter circular sheet of silicon carbide sandpaper, which has a transmissivity of about 40%. The SiC grains, about 40 μm in diameter, are opaque to x-rays, thus the beam intensity profile is modulated and a speckle pattern is transmitted, which is actually an x-ray absorption image projection. Since the beam diverges only slightly, to obtain a sufficiently large area of illumination to cover an actual sample the distance from the sandpaper to the object had to be 2.2 m, projecting an approximately 5×5 mm$^2$ square beam. The average size of the speckles on this plane was 0.4 mm. The sandpaper was mounted on a rotary motor fixed on a step motor, and was rotated through 360° in intervals of 0.4 to 0.9° for every 0.5 mm horizontal translation of the step motor to produce a series of random speckle patterns. A Photonic Science Large Area VHR22M_125 camera (CCD1, see Fig 1(a)) placed exactly in the object plane recorded each speckle pattern with an exposure time of 10 s. A total of more than 10,000 frames were recorded and stored for future use.

The camera was then removed and replaced by the object at the same position. Another x-ray camera (CCD2) was then placed 10 cm behind the object to measure the bucket signal, as shown in Fig 1(b); this distance is inconsequential so long as the entire beam can be collected. The sandpaper was rotated and translated in exactly the same sequence as before, and the total intensity transmitted through the object was integrated for each exposure after each step of the motor. A Princeton Instrument PIXIS-XB:1300 camera of 20 μm resolution was chosen to be the bucket detector CCD2 on account of its high sensitivity of about 50% quantum efficiency. The exposure time for each bucket signal was $t_0 = 1$ μs.

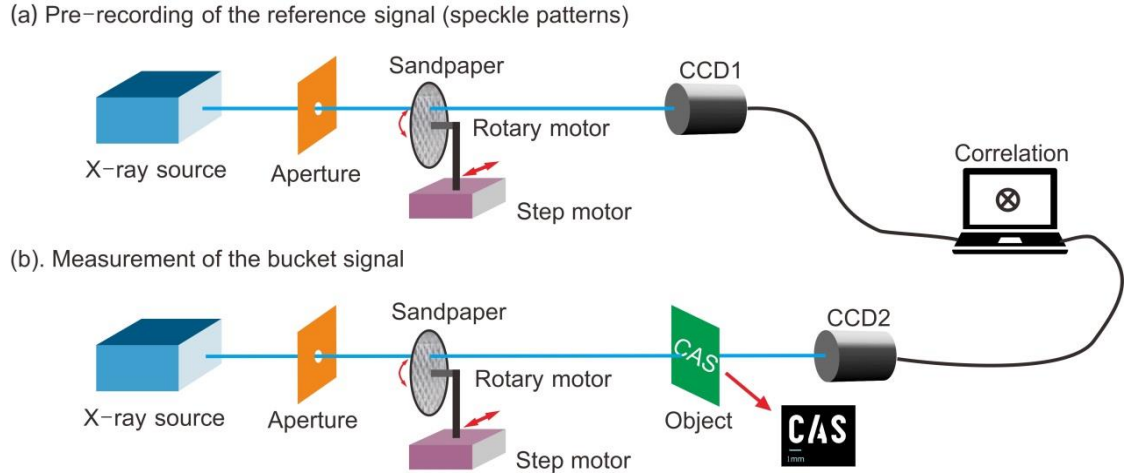

Fig 1. Experimental setup. (a) Pre-recording of the reference speckle patterns with CCD1; (b) Measurement of the field transmitted by the object with the bucket detector CCD2. The object and CCD1 are in the same plane.

Of course, it is important to check if the speckle pattern remains the same when the step motor returns to the same position every time, i.e. if the light field pre-recorded by the reference detector is identical to that during all subsequent exposures. This depends on the stability and uniformity of the x-ray source, as well as on the precision of the step motor. To check the uniformity of the output beam, we made a direct intensity measurement using CCD1; Fig. 2(a) shows a 3 s exposure shot of the beam profile. The plot looks rather noisy but we can see that there are noise fluctuations even outside the beam cross-section, indicating that the noise is due to the CCD electronics and not beam fluctuations (the zero intensity line is biased at 100 by the CCD software to compensate for this instrumental noise). The constancy of the speckle patterns was checked a couple of times, and Fig. 2(b) shows the first speckle pattern $I_1$ in the pre-recorded series of reference signals; Fig. 2(c) is the first speckle pattern $I_1'$ in the second series, used to illuminate the object. The two speckle patterns have slightly different gray scales because of the varying background noise of the CCD, but otherwise they are the same. This proves that our pre-recording scheme is practical and feasible for XGI.

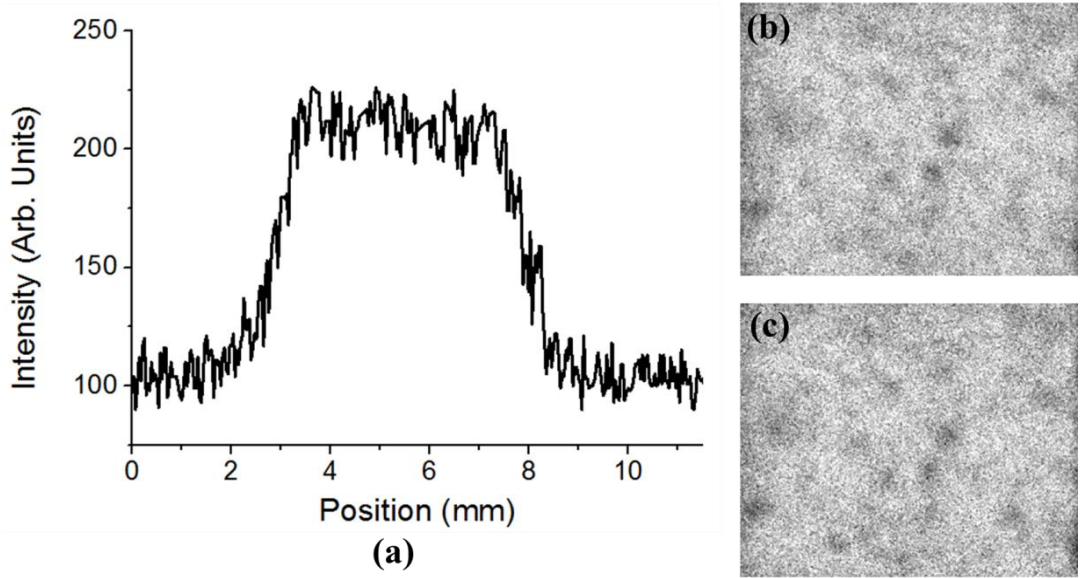

Fig. 2(a) Intensity profile of the direct x-ray beam; (b) Pre-recorded speckle pattern $I_1$; (c) Speckle pattern $I_1'$ in the second series of positions.

Finally, the ghost image is retrieved after *N* measurements by calculating the correlation

$$G(x,y) = \langle SI(x,y)\rangle - \langle S\rangle\langle I(x,y)\rangle \approx \frac{1}{N}\sum_{i=1}^{N} S_i I_i(x,y) - \frac{1}{N^2}\sum_{i=1}^{N} S_i \sum_{i=1}^{N} I_i(x,y),$$

where *x* and *y* are the horizontal and vertical pixel co-ordinates of the image, respectively, *S* is the integrated bucket detector intensity, and *I*(*x*, *y*) is the measured intensity at each pixel of CCD2.

The first object we imaged was a stainless steel mask with the stencilled letters "CAS", approximately 5 mm in width. The XGI image after averaging over $N = 10^4$ exposures is shown in Fig. 3(a). The total exposure time required for each XGI image is $t = Nt_0 = 10$ ms. Figure 3(b) shows a traditional projection x-ray (PX) image taken with CCD2 with the same exposure time of $t = 10$ ms; we can see that the image is barely discernible.

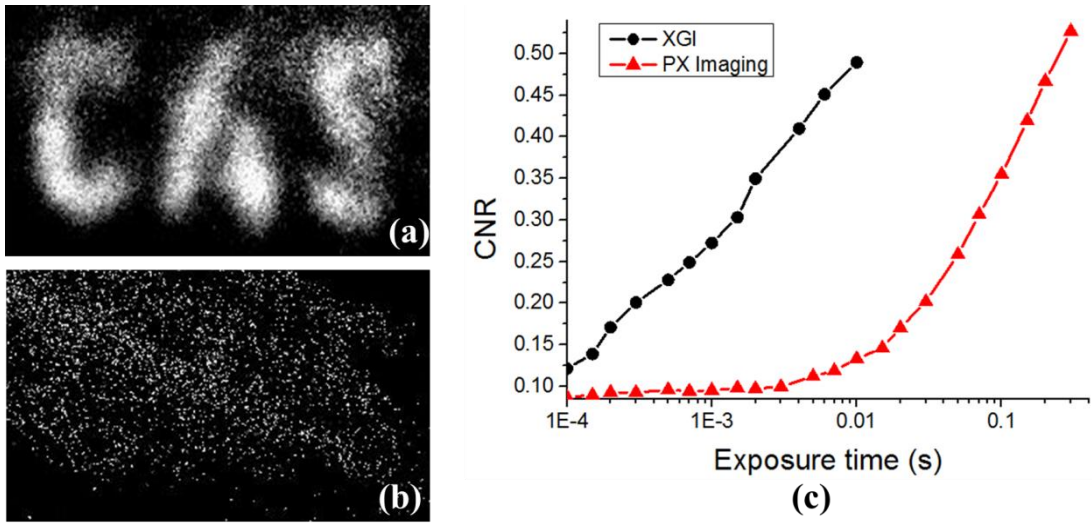

Fig. 3. Imaging results. (a) XGI image for $N = 10000$ frames and a total exposure time of 10 ms; the CNR is 0.5 (b) PX image after an exposure of 10 ms; the CNR is 0.13; (c) CNR vs. total exposure time.

The actual radiation dose in our XGI experiment was quantified using an imaging plate (Fuji Film, SR2025) placed in the beam after the sandpaper, and the total flux was estimated to be $2.9 \times 10^5$ photons/s.mm$^2$ [25]. This corresponds to 120 photons/s per pixel of CCD2, so since the exposure time for a complete ghost image was 10 ms, the total number of x-ray photons required to complete an image exposure by XGI is only 1.2 photons/pixel. This is significant for biological organisms when radiation damage is a concern.

An important quantitative indicator of image quality is the contrast-to-noise ratio (CNR), defined as [26]

$$CNR \equiv \frac{\langle G_{in}(\vec{r}) \rangle - \langle G_{out}(\vec{r}) \rangle}{\sqrt{\sigma_{in}^2 + \sigma_{out}^2}},$$

where $G_{in}$ and $G_{out}$ are the XGI signals for any point where the transmission is 1 and 0, respectively; $\sigma_{in}^2$ and $\sigma_{out}^2$ are the corresponding variances of the XGI signal, i.e. $\sigma^2 \equiv \langle G^2(\vec{r}) \rangle - \langle G(\vec{r}) \rangle^2$. The CNR of the image in Fig. 3(a) is calculated to be 0.5, while that of Fig. 3(b) is just 0.1.

A plot of the CNR of the images obtained by XGI and conventional PX imaging as a function of total exposure time is shown in Fig. 3(c); for XGI the abscissa values correspond to $Nt_0$, and it should be noted that this axis is logarithmic. We can see that for short exposure times, i.e. low radiation doses, the CNR of XGI (black circles) improves much more rapidly than PX imaging (red triangles).

To test XGI on a natural biological object, a small shell about 6 mm in length was chosen, see Fig. 4(a). Considering that larger objects may need to be imaged in the future, the large area CCD1 was chosen as the bucket detector, but since it is less sensitive than CCD2 the exposure time for each frame was lengthened to 220 ms. The XGI image is shown in Fig. 4(b), reconstructed from $10^4$ measurements. The image is obviously not as pretty as the ordinary photograph since it is a transmission image, but the contour is clear and the gray-scale regions contain rich information about the interior of the shell.

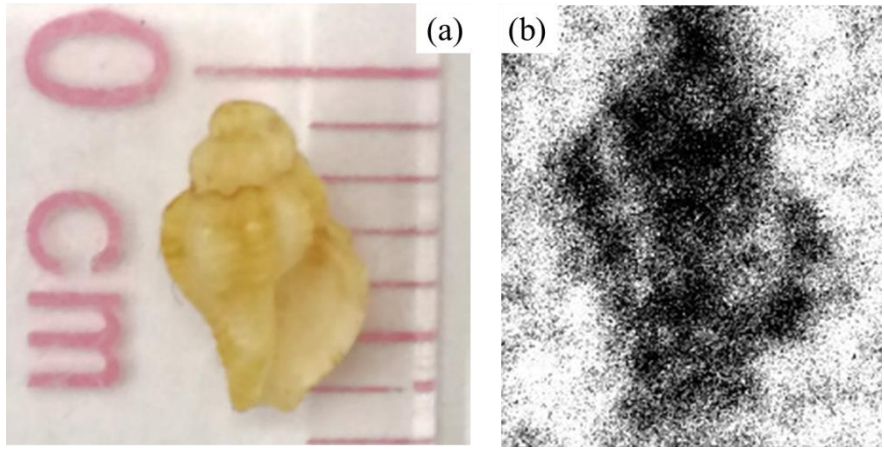

Fig. 4. (a) Ordinary photo of the shell; (b) XGI gray-scale transmission image.

It may be observed that the edges of the XGI images appear somewhat blurred. This is because in our experiment the resolution is determined by the speckle size of 0.4 mm, which is determined by the SiC grain size rather than the pixel size of the detector. If finer sandpaper were used, the resolution would be higher, and the edges would be much clearer.

In conclusion, we have realized XGI using a table-top source under ultra-low x-ray illumination. Real objects were imaged, with a quality surpassing that of conventional PX for the same low radiation dose. The setup is very simple, relatively inexpensive, and easy to operate. Pre-recording of the patterned illumination can be performed with a bright beam, while actual exposure of the object can be realized with a dosage on the order of single photons per pixel. This is obviously an important advantage in analysis of sensitive organisms or in vivo specimens. With finer speckle patterns or with various computational methods the spatial resolution could be further improved, while by shortening the longitudinal coherence length it should be possible to perform tomographical XGI.